\documentclass[%
    reprint,
    showpacs,preprintnumbers,
    showkeys,
    amsmath,amssymb,
    aps,
    prb,
    floatfix,
    longbibliography,
]{revtex4-2}

\usepackage[english]{babel}
\usepackage[utf8]{inputenc}
\usepackage{graphicx}
\usepackage{bm}
\usepackage{hyperref}
\usepackage{xcolor}
\usepackage{epsfig}
\usepackage{amsmath}
\usepackage{amsfonts}
\usepackage{amssymb}
\usepackage{amsthm}
\usepackage{dsfont}
\usepackage[all]{nowidow}
\usepackage{pifont}
\usepackage{algorithm}
\usepackage{algpseudocode}
\usepackage[capitalise]{cleveref}
\usepackage{tabularx}
\usepackage{tikz}
\usepackage{xcolor}
\usepackage{xspace}

\let\doi\undefined
\usepackage{doi}

\usepackage[letterspace=130]{microtype}
\definecolor{refColor}{HTML}{0376E9}
\definecolor{figColor}{HTML}{E90303}
\definecolor{urlColor}{HTML}{0376E9}

\hypersetup{
    unicode=false,                 % non-Latin characters in Acrobat’s bookmarks
    pdftoolbar=true,               % show Acrobat’s toolbar?
    pdfmenubar=true,               % show Acrobat’s menu?
    pdffitwindow=false,            % window fit to page when opened
    pdfstartview={FitH},           % fits the width of the page to the window
    pdftitle={},                   % title
    pdfauthor={Nicolai Lang},      % author
    pdfsubject={Quantum physics},  % subject of the document
    pdfcreator={Nicolai Lang},     % creator of the document
    pdfproducer={Nicolai Lang},    % producer of the document
    pdfkeywords={},                % list of keywords
    pdfnewwindow=true,             % links in new window
    colorlinks=true,               % false: boxed links; true: colored links
    linkcolor=figColor,            % color of internal links
    citecolor=refColor,            % color of links to bibliography
    filecolor=magenta,             % color of file links
    urlcolor=urlColor              % color of external links
}

\newcommand{\ket}[1]{\mathinner{|{#1}\rangle}}

\newcommand{\CaptionMark}[1]{\textbf{#1}}
\newcommand{\lbl}[1]{({#1})}

\newcommand{\note}[1]{{\color{blue}#1}}

\graphicspath{{./graphics/}}
\newcommand{\CondSX}{$\mathrm{Cond}\sqrt{X}$\xspace}
\newcommand{\CondSXX}{$\mathrm{Cond}\sqrt{XX}$\xspace}
\newcommand{\CondSXXX}{$\mathrm{Cond}\sqrt{XXX}$\xspace}
\newcommand{\CSX}{$\mathrm{C}\sqrt{X}$\xspace}
\newcommand{\CNOT}{$\mathrm{CNOT}$\xspace}

\begin{document}

\title{Absorbing State Phase Transition with Clifford Circuits}

\author{Nastasia Makki} \email{nastasia.makki@itp3.uni-stuttgart.de}
\author{Nicolai Lang}
\author{Hans Peter Büchler}

\affiliation{%
    Institute for Theoretical Physics III 
    and Center for Integrated Quantum Science and Technology,\\
    University of Stuttgart, 70550 Stuttgart, Germany
}

\date{\today}

%+++++++++++++++++++++++++++++++++++++++++++++++++++++++++++++++++++++++++++++++
%% ABSTRACT
%+++++++++++++++++++++++++++++++++++++++++++++++++++++++++++++++++++++++++++++++

\begin{abstract}
The role of quantum fluctuations in modifying the critical behavior of non-equilibrium
phase transitions is a fundamental but unsolved question. In this study, we examine
the absorbing state phase transition of a 1D chain of qubits undergoing a contact process
that involves both coherent and classical dynamics. We adopt a discrete-time quantum model
with states that can be described in the stabilizer formalism, and therefore allows for an efficient
simulation of large system sizes. The extracted critical exponents indicate that the absorbing state phase transition of this Clifford circuit model belongs to the directed percolation universality class. 
This suggests that the inclusion of quantum fluctuations does not necessarily alter the critical behavior of non-equilibrium phase transitions of purely classical systems. 
Finally, we extend our analysis to a non-Clifford circuit model, where a tentative scaling analysis in small systems reveals critical exponents that are also consistent with the directed percolation universality class.
\end{abstract}

\maketitle

%\todo{Replace all figures by vectorgraphics (PDF).}

%+++++++++++++++++++++++++++++++++++++++++++++++++++++++++++++++++++++++++++++++
\section{Introduction}
%+++++++++++++++++++++++++++++++++++++++++++++++++++++++++++++++++++++++++++++++
Non-equilibrium quantum phase transitions are interesting because they can exhibit
universal behavior that is distinct from classical systems \cite{sieberer2016keldysh,PhysRevLett.110.195301}, but are in general much harder to study. Recently,
quantum contact models have been proposed as promising systems to explore this possibility \cite{marcuzzi2016absorbing}. Classical
contact models are among the conceptually simplest to investigate
non-equilibrium phase transitions. They are characterized by the competition of a spreading mechanism and a probabilistic decay, and are widely utilized in describing phenomena
like the spread of diseases, forest fires and bacteria colonies~\cite{mollison1977spatial,grassberger1983critical,diekmann2000mathematical,sullivan2009wildland,bonachela2011universality}.
For a low spreading rate, the system reaches a state with no infected units and remains
struck in this absorbing state. However, as the spreading rate increases beyond a critical
value, the system reaches an active state where the density of infected units fluctuates
around a fixed finite value. The phase transition in classical contact processes is well
understood: it falls under the directed percolation (DP) universality class, and all critical exponents are known to high precision~\cite{hinrichsen2000non}. Quantum analogs of
contact processes, on the other hand, remain less well understood. Since the computational capacities of
classical computers limit the exact simulation of generic many-body quantum systems to small
sizes, many questions concerning their quantum critical behavior remain
unanswered. Here, we study the critical properties of an absorbing state phase transition in a random quantum circuit model based on Clifford operations, which can be efficiently studied numerically.

The study of quantum contact processes has been motivated by a potential experimental
realization with Rydberg atoms~\cite{PhysRevB.95.014308}. 
Such neutral atoms, excited to Rydberg states and individually trapped by optical tweezers
or in the frozen regime, have emerged as a highly promising platform for the quantum simulation
of equilibrium and non-equilibrium quantum many-body 
systems~\cite{weimer2010rydberg,RevModPhys.82.2313,zeiher2016many,bernien2017probing,vsibalic2018rydberg,PhysRevLett.120.180502,de2019observation,browaeys2020many,PhysRevX.11.031005,semeghini2021probing}.
The interplay of coherent driving, strong interactions between the Rydberg states,
and spontaneous decay also opens the possibility to explore models closely
related to quantum contact processes~\cite{wintermantel2021epidemic}. Convenient observables, such as population loss 
and excitation density, display power-law scaling with the driving strength, and the inferred
exponents provide insight into the critical behavior of such systems~\cite{helmrich2018uncovering}. First theoretical studies using a mean-field approach suggest a first order transition for the pure quantum
contact model, and including classical contributions leads to the existence of a bicritical point,
for which the universal behavior deviates from the classical continuous DP transition~\cite{marcuzzi2016absorbing,PhysRevB.95.014308}. A renormalization group analysis concludes that strong temporal
and spatial fluctuations in the active phase smooths
out the first order transition predicted by mean-field approximations~\cite{roscher2018phenomenology}. Numerical simulations of a 50 site
chain using a tensor network (iTEBD) algorithm report a continuous absorbing state phase transition and provide first estimates for the critical exponents of a new quantum contact universality class~\cite{gillman2019numerical,carollo2019critical,carollo2022nonequilibrium}. A machine learning approach, employed 
to pinpoint the critical region, followed by a tensor network and quantum jump Monte Carlo
analysis also provide estimates for the critical exponents~\cite{PhysRevResearch.3.013238}; they find that only one (decay) exponent differs from the DP value, and only in the
case where the system is initialized in a homogeneous, all active state. 
In general, numerical methods are strongly limited by the size of the system they allow to simulate~\cite{RevModPhys.93.015008}.
The finite bond dimension of tensor networks limits this method to systems with
low entanglement, another obstacle that makes accurate studies of critical quantum behavior challenging. However, there are alternative numerical methods based on stabilizer
states which allow for the exploration of quantum systems with extensive entanglement.  
Such Clifford quantum circuits can simulate unitary as well as dissipative operations, and can
be efficiently implemented on classical computers~\cite{PhysRevA.70.052328}. This approach
has recently been successfully applied to investigate entanglement transitions~\cite{PhysRevB.98.205136,PhysRevX.9.031009,PhysRevB.99.224307,PhysRevB.100.134306}, and, in particular, systems where an absorbing state transition occurs as well~\cite{PhysRevLett.130.120402,PhysRevB.107.224303}.

In this manuscript, we study a quantum circuit version of a contact process based on Clifford operations. 
As universal properties are independent of the microscopic realization, we expect a critical behavior that can also be realized in continuous time quantum contact models. 
At each discrete time step we apply with some probability a set of unitary gates to represent coherent time dynamics, and some projective measurements to simulate dissipative decay. 
More importantly, we restrict the gates to the Clifford group of unitaries, which restricts the Hilbert space and thus allows for the efficient simulation of the system near its thermodynamic limit. 
Our results show that contact process transitions simulated with Clifford circuits belong to the directed percolation universality class. 
We observe that keeping the absorbing state dynamically stable, while restricting the Hilbert space to stabilizer states, leads to steady states with limited entanglement at the phase transition as well as in the active phase. 
The quantum effects included in this approach are insufficient to change the universality class of the classical contact process. 
To check if abandoning the Clifford restriction causes a distinctive change to the dynamics, we introduce a non-Clifford quantum contact process and study its non-equilibrium properties on a small system of 20 qubits.
The data suggests a volume law entanglement scaling in the active phase of this non-Clifford quantum contact model. 
Because of the strong finite size effects, we cannot reliably extract critical exponents from the exact simulation data. 
However, if we assume for the critical exponents the values of DP, we find that the scaling behavior for small systems is still consistent with this assumption. 
These inconclusive results emphasize the importance of realizing such models on emerging quantum simulation platforms where larger systems can be analysed.

%+++++++++++++++++++++++++++++++++++++++++++++++++++++++++++++++++++++++++++++++
\section{Model}
%+++++++++++++++++++++++++++++++++++++++++++++++++++++++++++++++++++++++++++++++

%///////////////////////////////////////////////////////////////////////////////

%% Define pictograms for states and gates

\DeclareRobustCommand{\tzactive}{\ket{\tikz[baseline=-0.6ex]{%
    \fill[red] circle(3pt);%
}}}
\DeclareRobustCommand{\tzinactive}{\ket{\tikz[baseline=-0.6ex]{%
    \fill[gray] circle(3pt);%
}}}
\DeclareRobustCommand{\tzsuper}{\ket{\tikz[baseline=-0.6ex]{%
    \fill[green] circle(3pt);%
}}}
\DeclareRobustCommand{\tzpair}{\ket{\tikz[baseline=-0.6ex]{%
    \fill[green] (0,0) circle (3pt);
    \fill[green] (.5,0) circle(3pt);
    \draw[green, line width=2pt] (0,0) -- (0.5,0);
}}}
\DeclareRobustCommand{\tzdecay}{\tikz[baseline=.1ex]{%
    \fill[gray!60] rectangle (6pt,6pt);
}\xspace}
\DeclareRobustCommand{\tzcnot}{\tikz[baseline=-.6ex]{%
    \fill[black] (0,0) circle (2.5pt); 
    \node[circle, fill=white, draw=black, inner sep=0pt, line width=1pt] (A) at (0.5,0) {\footnotesize\textbf{+}};
    \draw[black, line width=0.5pt] (0,0) -- (A);
}\xspace}
\DeclareRobustCommand{\tzcsqrtx}{\tikz[baseline=-.6ex]{%
    \fill[black] (0,0) circle (2.5pt); 
    \node[circle, fill=white, draw=black, inner sep=0.5pt, line width=1pt] (A) at (0.5,0) {%
        \scalebox{0.8}{\tiny$\sqrt{X}$}};
    \draw[black, line width=0.5pt] (0,0) -- (A);
}\xspace}
\DeclareRobustCommand{\tzcondsqrtx}{\tikz[baseline=-.6ex]{%
    \fill[black] (-2pt,-2pt) rectangle (2pt,2pt); 
    \node[circle, fill=white, draw=black, inner sep=0.5pt, line width=1pt] (A) at (0.5,0) {%
        \scalebox{0.8}{\tiny$\sqrt{X}$}};
    \draw[black, line width=0.5pt] (0,0) -- (A);
}\xspace}

\begin{figure*}[tb]
    \centering
    \includegraphics[width=0.95\textwidth]{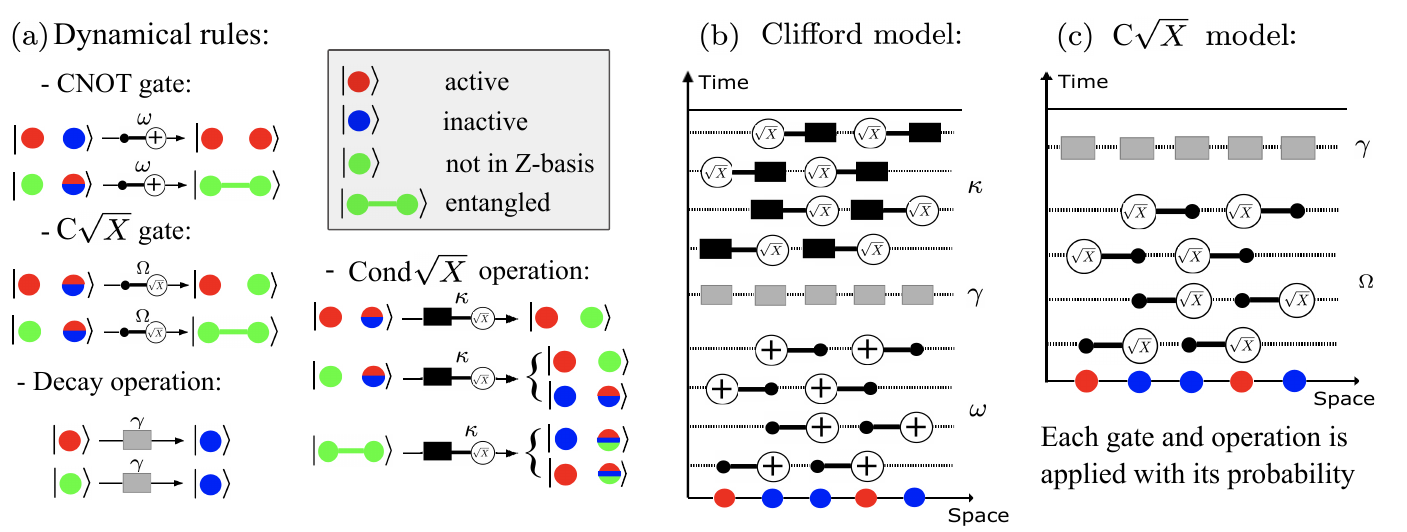}
    \caption{%
        \CaptionMark{Quantum contact models.} 
        The basis states $\tzactive=\ket{\mathrm{active}}$ and
        $\tzinactive=\ket{\mathrm{inactive}}$ are eigenstates of the
        Pauli operator~$Z$. $\tzsuper$ represents a site that is not in a
        $Z$-eigenstate and $\tzpair$ is a generic entangled state.
        \lbl{a}~The dynamical rules: First, the decay operation \tzdecay
        is applied with probability $\gamma$. It is a single site process where
        a projective Z-measurement is performed on the qubit followed by an X gate
        to flip the state if the measurement outcome was  $\ket{\mathrm{active}}$. 
        The \CNOT gate \tzcnot is applied with a probability $\omega$. The gate can 
        activate an inactive site if the control was active. An entangled state 
        is created if the control is in a superposition. The controlled-$\sqrt{X}$ 
        (\CSX) gate \tzcsqrtx is applied with a probability $\Omega$. A product
        state where the control is active leads to the creation of a local
        superposition on the target site. Another application of the gate
        activates an initially inactive target. Entangled pair states can
        be created as well if the control is in a superposition. Finally,
        we define a two-qubit operation: The conditioned-$\sqrt{X}$ (\CondSX)
        gate \tzcondsqrtx is applied with a probability $\kappa$. The condition
        is based on a measurement outcome of the control site; if active, the
        target is acted upon with a $\sqrt{X}$ gate. This process decoheres
        an entangled state, but creates a local superposition after being
        applied to a product state in the $Z$-basis.
        \lbl{b}~The Clifford model is defined as a quantum circuit that
        combines layers of \CNOT gates and \CondSX operations, applied
        symmetrically in a checker board manner, with a layer of single site
        decay operations. The sketch above represents a single time step
        where each operation is applied with a given probability. For a fixed
        finite decay probability $\gamma$, there exists a critical $\omega$
        and $\kappa$ bellow which the system evolves into a dynamically
        stable absorbing phase defined as all sites are inactive. 
        \lbl{c}~The \CSX model is defined as a quantum circuit built
        from layers of \CSX gates and decay operations. We expect
        to observe an absorbing state phase transition as we increase the
        probability $\Omega$ at constant decay $\gamma$ where the dynamics is
        dominated by quantum effects. However, this is a non-Clifford circuit
        which can be numerically simulated only for small system sizes.
    }
    \label{fig:model}
\end{figure*}
%///////////////////////////////////////////////////////////////////////////////

We focus on a 1D lattice of length $L$ with periodic boundaries. Each site is
a two-level quantum system represented by a qubit state. The basis states are
labeled $\left | 1  \right \rangle$ and $\left | 0  \right \rangle$ which
we translate to contact process language as $\left | \textnormal{active}
\right \rangle$ and $\left | \textnormal{inactive} \right \rangle$ states
respectively.
%\note{We start with a discretization of the time evolution
%in the models.} \todo{I find this misleading.} 
At each time step, a set of
dynamical rules apply which define how activation can spread and coagulate
within a system. The simplest classical contact process requires two dynamical
rules. First, sites can become active if and only if one of their nearest
neighbors is already in an active state, i.e., by contact. Second, active
sites can spontaneously decay and become inactive. Spontaneous activation
should be impossible which makes the state where all the sites are inactive
invariant under the dynamical rules. Therefore, once a system ends up in the
fully inactive state, it gets stuck in this state which is therefore termed
\textit{absorbing}. The other dynamically stable state of this system is
when the spreading and decay processes balance each other such the density
of active sites fluctuates around a finite fixed value. The system is
then said to be in an \textit{active phase}. The relative strength of the
spreading probability and decay probability determine the steady state and changing it drives a phase transition. The presence of an absorbing state
violates detailed balance, making this transition an out-of-equilibrium phase
transition. In a classical system, the sites can only exist in one of the two
basis states, either active or inactive. The classical phase transition is
well studied and classified into the directed percolation (DP) universality
class; for a detailed review on the classical contact process we refer the
reader to~\cite{hinrichsen2000non}.

To investigate quantum analogs of this phase transition, we map the problem
to a discrete time quantum circuit where unitary gates and measurements are
chosen to reproduce the dynamical rules of a contact process applied now to
a chain of qubits. In this paper, we study two models that implement the
ideas of the contact process in a quantum system. The first is a Clifford
model where we restrict the gates to the set of Clifford unitaries: Hadamard
($\textrm{H}$), Controlled-Not (\CNOT), and Phase ($\textrm{P}$) gates. Operations that can be 
written in terms of the Clifford set are also included such as the Pauli 
matrices: $\textrm{X}$, $\textrm{Y}$ and $\textrm{Z}$. In the second model, we relax this restriction. 
\cref{fig:model} shows the dynamical rules applied
by the different activation spreading operations in our two models. The
decay operation is the same for both models; it is achieved via a projective
single site measurement onto the $\textrm{Z}$-basis followed by a flip using the $\textrm{X}$ gate if the measurement result was an active site. This operation is applied on each site with a constant probability $\gamma$. 

In the Clifford model, activation is spread via layers of \CNOT gates,
each applied with a probability $\omega$, and via layers of \CondSX operations, 
each applied with a probability $\kappa$. The conditioned-$\sqrt{X}$ operation 
is defined to first \textit{measure} the control qubit and only if the measurement yields the control qubit in the active state, apply a $\sqrt{X}$ gate on the target qubit.
We start with the simplest Clifford contact model where $\kappa=0$. 
This model is referred to as the \CNOT model. At a fixed
decay rate $\gamma=0.1$, varying the spreading coupling $\omega$ drives
the system through an absorbing state phase transition. The analysis of the
critical point is shown in \cref{figclassical} (\cref{app:scaling}), where
the scaling behavior is found to fall into the directed percolation (DP)
universality class. The equivalence of the \CNOT model to a classical contact
model can be easily seen if one considers initializing the system in a product
state of sites in the $\textrm{Z}$-basis, i.e., a classical state. The dynamical rules
dictate that the system remains in such a classical state as entangled states
can be created only if the control site of the applied \CNOT gate happens
to be in a superposition of the basis states. Therefore, to observe a finite
entanglement entropy of a chosen subsystem with the rest, we should start our
simulations with a state were some sites are not in the $\textrm{Z}$-basis. However,
the decay operation continuously removes entanglement and we find that for any
finite $\gamma$, the system eventually converges to a classical product state
where sites are either active or inactive. A similar result is obtained if
one considers the Clifford model with $\omega=0$ and $\kappa >0$ which we refer to as the \CondSX model. 
Increasing $\kappa$ at a fixed decay rate also drives the system through 
an absorbing state phase transition within the DP universality class, 
as shown in \cref{figclassical2} in \cref{app:scaling}. The critical point 
occurs at twice the value of the \CNOT model which is explained by the fact 
that for classical states, two conditioned-$\sqrt{X}$ operations are equivalent 
to a \CNOT gate. 
We expect that the character of the phase transition can change 
in a quantum circuit where dynamical creation of entanglement entropy is possible, where gates that rotate the local state and not just flip it are used to spread activation. However, if we replace the \CNOT with a controlled-Hadamard ($\mathrm{CH}$) or a controlled-$\sqrt{X}$ 
(\CSX), the resulting contact models are non-Clifford and thus cannot be simulated 
efficiently on classical computers. In \cref{sec:nonClifford} we present an analysis of 
the absorbing state phase transition that occurs in the \CSX model, depicted in 
\cref{fig:model}c, for a small chain of 20 qubits. In the following, we focus on the 
Clifford model shown in \cref{fig:model}b. When both probabilities $\kappa$ and $\omega$ 
are finite, entangled states can be created dynamically but require two steps. First,
a \CondSX operation is needed to create a local superposition and
then subsequent \CNOT gates can spread the entanglement.

%+++++++++++++++++++++++++++++++++++++++++++++++++++++++++++++++++++++++++++++++
\section{Results}
%+++++++++++++++++++++++++++++++++++++++++++++++++++++++++++++++++++++++++++++++

%///////////////////////////////////////////////////////////////////////////////
\begin{figure}[tb]
    \centering
    \includegraphics[width=\columnwidth]{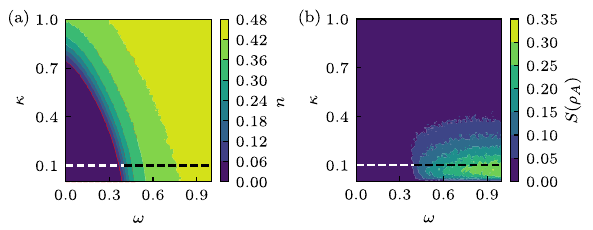}
     \caption{% 
         \CaptionMark{Phase diagram of the Clifford model.} 
         \lbl{a}~For a fixed decay constant ($\gamma=0.1$) we observe an
         absorbing phase with zero steady state density in the region of small
         spreading parameters $\omega\lesssim 0.41$ and $\kappa\lesssim 0.79$.
         \lbl{b}~Shows the corresponding steady state half-system entanglement
         entropy. In the absorbing phase and along the axis $\omega=0$ and
         $\kappa=0$, the steady state is a product state so $S(\rho_A)=0$. The
         entropy decreases with increasing rate of projective measurements
         given by $\kappa$. We focus on the density transition in the region
         of small but finite $\kappa$ where we observe a finite entanglement
         in the active phase.
     }
    \label{Result1}
\end{figure}
%///////////////////////////////////////////////////////////////////////////////

\begin{figure}[tb]
    \centering
    \includegraphics[width=\columnwidth]{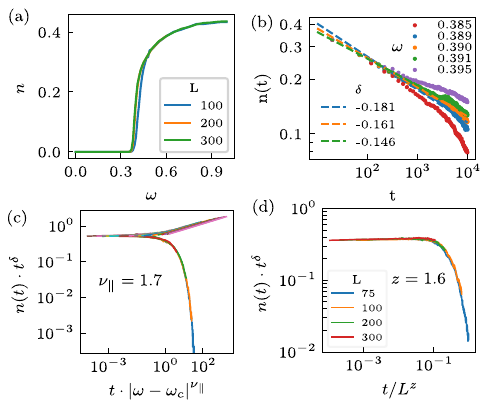}
    \caption{% 
        \CaptionMark{Scaling analysis of the Clifford model.}
        \lbl{a}~At fixed $\gamma=\kappa=0.1$, the order parameter $n(t_s)$, the
        steady state density of active sites, increases continuously from
        zero as a function of $\omega$. 
        \lbl{b}~The critical point of this second order transition is estimated
        around $\omega_c=0.390$ where $n(t)$ decays as a power as a function
        of simulation time. We then deduce the value of the decay exponent
        $\delta=0.161(9)$ from the slope of the fitted straight line in the
        double logarithmic scale.
        \lbl{c}~The best data collapse is obtained for the temporal correlation
        exponent $\nu_\parallel=1.70(3)$, such that $\beta=\delta \cdot
        \nu_\parallel=0.27(2)$.
        \lbl{d}~Also, near criticality, the data from different
        sizes collapses into a universal scaling function for
        $z=1.60(3)=\nu_\parallel / \nu_\perp$, so the spacial correlation
        exponent is $\nu_\perp=1.06(4)$. All derived exponents are in good
        agreement with the DP universality class (\cref{tab:exp}).
    }
    \label{Result2}
\end{figure}

Here we present the results for the Clifford model simulation. We initialize the 
system with half the sites activated, chosen randomly for each realization. The
initial state is then evolved by applying the quantum circuit for $t_s$
time steps; this is repeated for $10^3$ realizations. The main observable
is the average density of active sites $n(t)$ as a function of time $t$;
this quantity depends on the initial state on short timescales but later
acquires a behavior independent of the initial condition. The number of applied 
time steps $t_s$ for each simulation should be chosen sufficiently long for 
the system to reach the steady state regime. The density $n$ of
active sites in the steady state is the order parameter of the absorbing state 
phase transition. It changes from zero in the absorbing phase to a finite value 
in the active phase. 
It is important to note that in a finite size simulation, the steady state in the 
active phase is only quasi-steady; an initial state which evolved into a quasi-steady 
state remains there for a relatively long time compared to the time needed for 
equilibration, before eventually decaying into the absorbing state. The lifetime 
of a quasi-steady state grows exponentially with the size of the considered system. 
In our simulations, system sizes up to $L=400$ sites and $t_s=10^{4}$ time steps
are considered. \cref{Result1}a shows the phase diagram obtained for the Clifford 
model at constant decay rate $\gamma=0.1$, size $L=200$ and a steady state time of 
$t_s=10^2$ steps. The absorbing phase exists for small values of the two spreading 
couplings $\omega$ and $\kappa$, representing the coherent and incoherent dynamics 
respectively. The corresponding steady state entanglement entropy $S(\rho_A)$, 
where $\rho_A$ denotes the subsystem of size $L_A=L/2$, is plotted in \cref{Result1}b. 
It vanishes in the absorbing phase where all sites are inactive, also in regions where 
the dynamics ends up in a product state, like on the $\kappa=0$ and $\omega=0$ axes where states are classical by construction of the model. $S(\rho_{A})$ continuously decreases with increasing $\kappa$, as each \CondSX operation includes a projective measurement. Therefore, the interesting regime to study the phase
transition appears in the region of small $\kappa$ where $S(\rho_A)$
is finite in the active phase. We consider the absorbing phase transition along
the highlighted line $\kappa=\gamma=0.1$. \cref{Result2}a shows the change in the order 
parameter $n$ as a function of the \CNOT rate $\omega$. The phase transition is clearly
continuous. The scaling analysis, which estimates the critical exponents,
shows that all values agree with the known directed percolation exponents
that appear in the classical contact process, \cref{Result2}b-d; see
\cref{app:scaling} for details on the scaling analysis.

%///////////////////////////////////////////////////////////////////////////////
%///////////////////////////////////////////////////////////////////////////////

This result shows that the classical universality class can persists in a
quantum model with dynamically generated entanglement. However, in
our Clifford model, the entanglement entropy in the active phase is small
and remains in the area law scaling regime.

%///////////////////////////////////////////////////////////////////////////////
\begin{figure}[tb]
    \centering
    \includegraphics[width=\columnwidth]{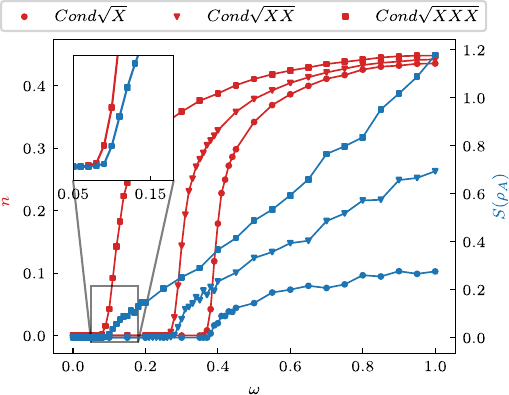}
    \caption{%
        \CaptionMark{Increased range of contact.} 
        We replace the \CondSX operation in our Clifford model first with
        \CondSXX and then \CondSXXX. Simulating the dynamics at the same fixed
        values for the probabilities $\gamma=\kappa=0.1$, we observe an increase 
        in the values of entanglement entropy reached in the active phase. The nature of
        the transition to the absorbing phase remains continuous and within
        the DP universality class. Note that the entanglement also grows
        continuously from zero and remains small at the transition.
    }
    \label{Result5}
\end{figure}
%///////////////////////////////////////////////////////////////////////////////

One can try to create more entanglement by increasing the relative effect of
creating local superpositions (versus projective measurements) and applying
more than one $\sqrt{X}$ gate per measurement. We refer to these multi-qubit
operations as \CondSXX and \CondSXXX. Such operations can be interpreted in the
context of the contact process as an increase in the range of interaction. In
\cref{Result5} we compare the absorbing state transitions obtained in these
models. We observe that despite reaching considerably higher values for the
half-system entanglement entropy, the nature of the phase transition remains
clearly within the directed percolation universality class. The critical point
occurs for lower values of the coupling $\omega_c$, which is expected. The
scaling analysis is given in \cref{fignext} in \cref{app:range}.

We note that in all these cases the entanglement entropy grows from zero
continuously when crossing the transition into the active phase and seems that
at the critical point, it remains too low to affect the universal behavior.

\begin{table}[tb]
    \centering
    \begin{tabularx}{\linewidth}{XXXXX} 
        \hline
        & DP & \CNOT & Clifford & \CondSXXX \\ 
        \hline
        $\delta$ &0.1595 &0.161(6) &0.161(9)  &0.154(6) \\
        $\beta$ &0.2765 &0.27(2)  &0.27(2)  &0.26(2) \\
        $\nu_{\parallel}$ &1.7338 &1.70(3)  &1.70(3)  &1.70(3)  \\
        $\nu_{\perp}$ &1.0969 &1.06(4)  &1.06(4)  &1.06(4)  \\ 
        \hline
    \end{tabularx}
    \caption{%
        \CaptionMark{Critical exponents.} 
        The critical behavior of all contact processes within
        the Clifford formalism is consistent with DP.
    }
    \label{tab:exp}
\end{table}

%+++++++++++++++++++++++++++++++++++++++++++++++++++++++++++++++++++++++++++++++
\section{Non-Clifford Model}
%+++++++++++++++++++++++++++++++++++++++++++++++++++++++++++++++++++++++++++++++
\label{sec:nonClifford}

\begin{figure}[tb]
    \centering
    \includegraphics[width=\columnwidth]{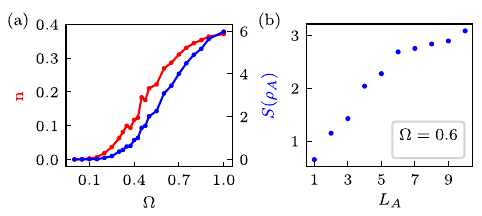}
    \caption{%
        \CaptionMark{The \CSX model.} 
        \lbl{a}~For a small system of size $L=20$ qubits, exact simulations 
        indicate a continuous absorbing state transition where both, the
        steady state density and the entanglement entropy increase continuously
        from zero with rate $\Omega$ at fixed decay rate $\gamma=0.1$.
        \lbl{b}~For $\Omega=0.6$, where the system appears to be in the
        active phase, the entanglement entropy scales with the volume of the
        subsystem: Plotting $S(\rho_A)$ versus the size $L_A$ of subsystem
        $A$ shows a linear growth when $L_A$ is small compared to the full
        system size. The increase then saturates due to the finite size.
    }
    \label{Result6}
\end{figure}

Now we focus on the non-Clifford \CSX model. The \CSX gate is unitary and 
results in a coherent spreading dynamics and the dynamical creation of entangled states. 
For a relatively weak rate $\Omega$ of applying \CSX gates, a finite decay rate 
drives the system into a classical absorbing phase, the existence of which is guaranteed  
by a finite threshold for anisotropic bond percolation on the square lattice \cite{Redner_1979} 
(here: the space-time lattice in \cref{fig:model}c).
By increasing $\Omega$, a transition from the absorbing phase into an active phase is expected. 
\cref{Result6}a shows the variation in the normalized density of active sites $n$ as a 
function of $\Omega$ at constant $\gamma=0.1$ and the corresponding entanglement entropy
$S(\rho_A)$ obtained after $t=100$ steps on a small system of 20 qubits. Simulating for 
longer times would result in a decay of $n$ driven by finite size effects, so the steady
state regime is hard to capture in such small system sizes. Nevertheless, we observe a key
difference with respect to the Clifford models. The entanglement entropy $S(\rho_A)$ in 
the active phase reaches much higher values. It seems to scale as a volume law, indicated 
by the linear increase of $S(\rho_A)$ as a function of subsystem size $L_A$ especially 
for small $L_A$ relative to $L$ as shown in \cref{Result6}b. On the other hand, and similar to
the behavior observed for the Clifford model transition, we find that both $n$ and $S(\rho_A)$ 
increase continuously from zero to finite values as we increase $\Omega$, which is consistent 
with a second order phase transition.

%///////////////////////////////////////////////////////////////////////////////
%///////////////////////////////////////////////////////////////////////////////

%///////////////////////////////////////////////////////////////////////////////
\begin{figure}[tb]
    \centering
    \includegraphics[width=0.48\textwidth]{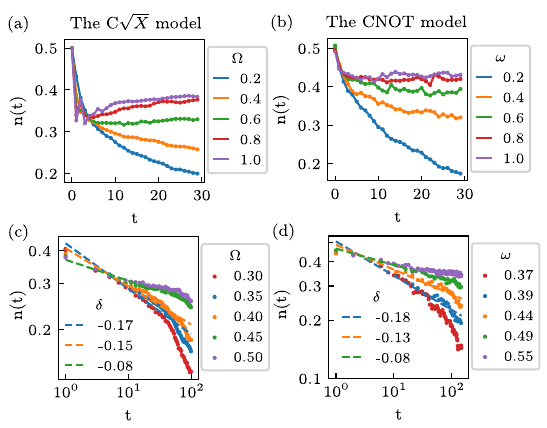}
    \caption{%
        \CaptionMark{Comparison of the \CSX and \CNOT models.} 
        \lbl{a,b}~Early time evolution of the order parameter $n$ in the
        \CSX and \CNOT models, respectively.
        \lbl{c,d}~Evolution of $n$ at later times. From larger systems,
        we know the critical point $\omega_c=0.39$ of the \CNOT model; the
        fitted power law at this point has a $10\%$ error with respect to
        the expected DP value. We can identify a range for $\omega_c$ where
        the decay in $n$ is linear in logarithmic scale up to  $t=100$,
        after which finite size effects dominate the dynamics. In the \CSX
        model, at short times we observe fluctuations in $n$ for larger
        $\Omega$. Another distinctive feature of the \CSX model is that $n(t)$
        smooths out to a lower average value compared to that of the \CNOT
        transition at similar coupling strength. Also finite size effects
        (in the form of overall decay of $n$) occur earlier than $t=100$
        steps. Despite the clear differences, we cannot exclude the DP value
        for the decay exponent $\delta$.
    }
    \label{Result7}
\end{figure}
%///////////////////////////////////////////////////////////////////////////////

To investigate the finite size effects on this transition, we compare the scaling 
behavior to that of the \CNOT model at an equal size $L=20$ but where the universal exponents are all well known, see \cref{Result7}. The finite size limits the time 
range for which data can be considered for the scaling analysis. Extracting critical exponents from this data is impossible, we can only check for its consistency with the known DP exponents.

In the case of the \CNOT model, the expected critical point $\omega_c$ and the decay exponent \note{$\delta$} both fall within the possible critical range observed
in the $L=20$ simulation data of \cref{Result7}d. In \cref{collCNOT} (\cref{app:small})
we show data collapse plots for various choices of fitted exponents as well as for the theoretical values of DP exponents while varying the critical point within this range. 
The quality of this data collapse into the universal scaling functions is consistently better for parameters closer to the results expected from simulating larger chains. Therefore we can conclude that for the \CNOT model the $L=20$ data fits well with the expected classical DP critical behavior up to simulation times around $t=100$ after which finite size effects become dominant.

\begin{figure}[tb]
    \centering
    \includegraphics[width=0.48\textwidth]{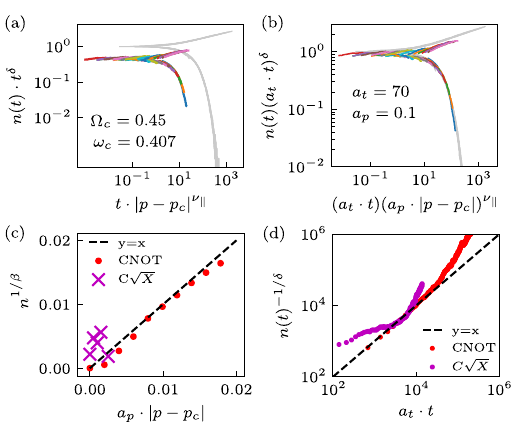}
    \caption{%
        \CaptionMark{Consistency with a DP transition.} 
        \lbl{a}~At $\Omega_c=0.45$ we observe the best data collapse in
        the \CSX model using the known DP critical exponents. The
        universal DP scaling function is plotted in gray as a reference. 
        \lbl{b}~We scale the data using non-universal factors $a_t$ and $a_p$
        such that the scaling functions overlap.
        In \lbl{c} and \lbl{d} we check whether at the chosen critical
        point the data scales as expected with the estimated non-universal
        factors. For reference we plot the scaled $L=20$ data from
        the \CNOT model using non-universal factors derived from the
        thermodynamic $L=400$ system as shown in \cref{non-universal}
        (\cref{app:scaling}). For the $L=20$ systems, the data from the
        \CNOT model fits nicely up to large $t$ when finite size effects
        dominate. However, the data from the \CSX model doesn't fit the
        expected scaling behavior.
    }
    \label{resultquestionable}
\end{figure}

A similar analysis for the \CSX model reveals a few notable differences:
First, at short times $t<10$, the density of active sites $n$ fluctuates, an effect mainly observed for simulations with relatively large $\Omega$ where the system is expected to
end up in an active phase. Also finite size effects seem to be stronger, the overall decay in $n$ starts at earlier simulation time, as seen \cref{Result7}c. The scaling of the data is analyzed within a range of $\Omega$ which possibly include the critical value and are 
shown in \cref{collCSX} (\cref{app:small}). The collapse is of worse quality compared to the plots obtained with the \CNOT model for the same size. If we assume the values of the exponents $\beta$ and $\nu_{\parallel}$ to be the literature DP values, we observe the 
best collapse for $\Omega_c \approx 0.45$. In \cref{resultquestionable}a we plot the collapsed data from \CSX model with the universal collapse function in grey as a reference. \cref{resultquestionable}b shows that there exists a choice for non-universal factors $a_p$ and $a_t$ that leads to data falling on the universal DP scaling function. Non-universal factors are extracted from scaling data at the critical point (see \cref{non-universal}
in \cref{app:scaling}) which we can not do in the \CSX model. \cref{resultquestionable}c and \cref{resultquestionable}d show that the in the \CNOT model the data agrees with the expected scaling at the critical point despite the clear finite size effects at larger times, where as in the \CSX model the data scaling reveals an inconsistency. As a result, we conclude that the nature of the absorbing state transition in the \CSX model remains inconclusive.

%///////////////////////////////////////////////////////////////////////////////
%///////////////////////////////////////////////////////////////////////////////

%+++++++++++++++++++++++++++++++++++++++++++++++++++++++++++++++++++++++++++++++
\section{Summary and outlook}
%+++++++++++++++++++++++++++++++++++++++++++++++++++++++++++++++++++++++++++++++

We introduced two models of random quantum circuits, comprised of unitary
entangling gates and projective measurements, to study the fate of absorbing
state phase transitions of classical contact models in the presence of
entanglement. The common features of both models are the existence of an
absorbing state that is invariant under the time evolution, and a spreading
mechanism that allows for the proliferation of excitations in the systems
-- both necessary ingredients for the occurrence of an absorbing state
phase transition. Most importantly, both models feature mechanisms for the
\emph{dynamical generation of entanglement}, which distinguishes them from
classical contact processes.

The first model -- the \emph{Clifford model} -- makes use of Clifford
gates only and therefore allows for efficient numerical simulations within
the stabilizer formalism. With our extensive simulations, we identified an
out-of-equilibrium phase transition between an absorbing phase and an active
phase. The latter featured low but finite levels of entanglement in certain
parameter regimes, which qualifies our model as a \emph{quantum} version of
classical contact processes. A scaling analysis revealed critical exponents
that match the directed percolation universality class perfectly. Since
the latter also describes the critical behavior of \emph{classical} contact
processes, we can conclude that this universality class is robust against
certain types of quantum fluctuations.

Since Clifford unitaries only allow for the exploration of a measure-zero
subset of the full many-body Hilbert space, their dynamics might be too
restrictive to modify this universality class. Our second model -- the \CSX
model -- makes use of the controlled-$\sqrt{X}$ gate to implement both a
spreading mechanism and dynamical entanglement generation. This gate is
\emph{not} in the Clifford group, and we had to resort to exact evaluations
of the time evolution for small system sizes. The simulations revealed
signatures of an absorbing state phase transition with significantly stronger
entanglement in the active phase. A tentative scaling analysis revealed
critical exponents consistent with the directed percolation universality
class. However, because of the small system sizes and strong finite-size
effects, our results do not allow for a conclusive characterization of the
critical behavior of this process.

%It would be interesting to apply approximate numerical methods to non-Clifford
%models -- like the \CSX model -- to study their out-of-equilibrium behavior for
%it would be interesting to study potential restrictions imposed by Clifford
%unitaries on the critical behavior of quantum contact processes. For example,
%it may be possible to show that no quantum contact process based on Clifford
%unitaries alone can modify the universality class of its classical counterpart.

Quantum simulations of such non-Clifford models for large systems might be an interesting application for near-term quantum computing platforms.
The study of universal features is promising because these are expected to be independent of microscopic details, and therefore might tolerate certain types of noise and gate infidelities of NISQ devices.
On the theoretical side, it is an interesting question whether Clifford circuits in general are too restrictive to describe non-classical critical behavior, 
and whether the numerical study of quantum criticality in non-equilibrium settings is an inherently hard problem.

%+++++++++++++++++++++++++++++++++++++++++++++++++++++++++++++++++++++++++++++++
%% ACKNOWLEDGEMENTS
%+++++++++++++++++++++++++++++++++++++++++++++++++++++++++++++++++++++++++++++++

\begin{acknowledgments}
This project has received funding from the German Federal Ministry of Education and Research (BMBF) under 
the grants QRydDemo and MUNIQC-Atoms, as well as Horizon Europe programme HORIZON-CL4-2021-DIGITAL-EMERGING-01-30 via the project 101070144 (EuRyQa).

\end{acknowledgments}

%+++++++++++++++++++++++++++++++++++++++++++++++++++++++++++++++++++++++++++++++
%% REFERENCES
%+++++++++++++++++++++++++++++++++++++++++++++++++++++++++++++++++++++++++++++++

% \bibliographystyle{plain}
\bibliographystyle{bibstyle}
%\bibliography{bibliography}

\clearpage

%+++++++++++++++++++++++++++++++++++++++++++++++++++++++++++++++++++++++++++++++
%% APPENDIX
%+++++++++++++++++++++++++++++++++++++++++++++++++++++++++++++++++++++++++++++++
\begin{appendix}

%+++++++++++++++++++++++++++++++++++++++++++++++++++++++++++++++++++++++++++++++
\section{Scaling theory}
\label{app:scaling}
%+++++++++++++++++++++++++++++++++++++++++++++++++++++++++++++++++++++++++++++++

Here we discuss the scaling analysis, based on Ref.~\cite{hinrichsen2000non},
which describes how we estimate the critical exponents. Absorbing states phase
transitions studied in this paper are continuous, second order transitions
which are often characterized by universal scaling laws. Different physical
systems that exhibit equal sets of critical exponents and coinciding scaling
functions belong to the same universality class. A universality class is
insensitive to microscopic details of its systems, and usually depends only
on properties like dimension, range of interactions, and symmetries. The
same picture exists in out-of-equilibrium processes. Most notably, the
directed percolation (DP) universality class is labeled by the triplet
($\beta$, $\nu_\parallel$, $\nu_\perp$) of critical exponents; all other
exponents can be deduced from this set using scaling relations. Remark that,
in contrast to equilibrium critical phenomena, the dimension of ``time''
has a different character than the ``space'' dimension, and we distinguish
these by using the indices $\parallel$ for time and $\perp$ for space.
Contact processes produce an absorbing state phase transition which belong
to the DP universality class. The order parameter $n(t_s)$ is the normalized
steady state density of active sites
\begin{equation*}
    n(t_s)= \left \langle \frac{1}{L}\sum_i n_i(t_s)  \right \rangle\,,
\end{equation*}
where $\langle...\rangle$ is an ensemble average over many realizations,
$L$ is the total number of sites in the system, and $n_i(t)$ is the local
occupation at site $i$ at time $t$. In the thermodynamic limit and close to
the transition, the steady state density obeys the scaling
\begin{equation*}
n(t_s) \sim |p - p_c|^\beta\,,
\end{equation*}
where $p$ is the driving parameter of the transition and $p_c$ is its critical
value. The exponent $\beta$ is conventionally associated to the scaling of
the order parameter, which is the density of active sites in contact process
transitions. The spatial correlation length $\xi_\perp$ and the temporal
correlation length $\xi_\parallel$ also diverge near criticality with the
scaling laws
\begin{equation*}
    \xi_\perp \sim  | p -p_c  |^{-\nu_\perp} 
    \quad\text{and}\quad
    \quad \xi_\parallel\sim  | p -p_c  |^{-\nu_\parallel} ,
\end{equation*}
where the length scales are related through the dynamical critical exponent
$z=\nu_\parallel/\nu_\perp$, defined as $\xi_\parallel\sim\xi_\perp^z$. Those
length scales can be determined from their intuitive physical
interpretations. For example, $\xi_\parallel$ represents the average decay
time of clusters that spread from an initial seed in the absorbing phase,
while $\xi_\perp$ represents the average spatial width of such clusters.
Although the set ($\beta$, $\nu_\parallel$, $\nu_\perp$) can be already
estimated from the relations above by plotting the associated quantities
($n(t_s)$,$\xi_\perp$, $\xi_\parallel$) in the double logarithmic scale as
a function of $\Delta=|p-p_c|$ and extracting the slope of the resulting
straight lines, this estimate is known to be quite inaccurate since the
equilibration time to reach the stationary state grows rapidly as the critical
point is approached, an effect known as \textit{critical slowing down}. A
more accurate approach to extract the critical exponents is to plot the
universal scaling functions instead.

Consider the density scaling relation for a finite size system with $N=L^d$
sites where $d$ is the spatial dimension:
\begin{equation*}
   n(t) \sim t^{-\delta} f( \Delta \cdot t^{1/\nu_\parallel}, t^{d/z}/N)\,.
\end{equation*}
$f$ has the same functional form for all phase transitions in the DP
universality class and can depend only on scale-invariant ratios. Note
that a scaling transformation $x \rightarrow \Lambda x$ of lengths $x$ is
accompanied by rescaling of $t \rightarrow \Lambda^z t$, $\Delta \rightarrow
\Lambda^{-1/\nu_\perp} \Delta$ and $ n \rightarrow \Lambda^{-\beta/\nu_\perp}
n$. In the thermodynamic limit $N\rightarrow \infty$, the universal function $f$ 
tends to a constant. Therefore, the exponent $\delta$ describes the power law 
decay of the order parameter near criticality $\Delta \rightarrow 0$.
At large arguments $\zeta$, $f(\zeta) \rightarrow \zeta^{\delta
\cdot \nu_\parallel}$. Therefore in the steady state, as $t\rightarrow\infty$ 
we find the scaling relation
\begin{equation*}
    \beta=\delta \cdot \nu_\parallel\,.
\end{equation*}

%///////////////////////////////////////////////////////////////////////////////
\begin{figure}[tb]
    \centering
    \includegraphics[width=\columnwidth]{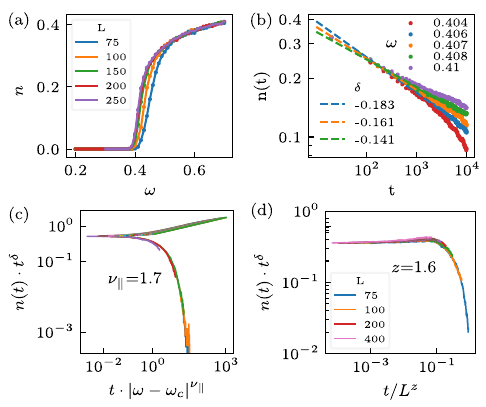}
    \caption{%
        \CaptionMark{The \CNOT model.} 
        \lbl{a}~Continuous change in the steady state density $n$ as a
        function of $\omega$ and fixed decay rate $\gamma=0.1$ for different
        system sizes. 
        \lbl{b}~Power law decay of the order parameter $n$ as a function
        of the simulation time. The critical point is estimated
        at $\omega_c=0.4070(5)$, the slope of the fitted line is
        $0.16087$ which gives the estimate for the critical exponent
        $\delta=0.161(6)$ (consistent with the best known estimate
        $\delta_\mathrm{DP}=0.159$~\cite{hinrichsen2000non}). The other
        critical exponents are deduced from the collapse plots \lbl{c}
        and \lbl{d}. All values are compatible with the literature values
        of the DP transitions.
    }
    \label{figclassical}
\end{figure}
%///////////////////////////////////////////////////////////////////////////////

Now we apply this analysis to the data obtained from the simulations of the
\CNOT model. We first plot the order parameter $n(t_s)$ as a function of
the rate $\omega$ for increasing system sizes and illustrate the continuous
nature of the phase transition in \cref{figclassical}a. Then we identify the
critical point using the expected asymptotic power law decay of the order
parameter $n(t) \sim t^{-\delta}$ for a sufficiently large system of $L=400$
sites and long simulation times $t_s=10^4$. Plotting the data in the double
logarithmic scale results in a positive (negative) curvature at large t for
coupling $\omega$ corresponding to the active (absorbing) phase as shown in
\cref{figclassical}b. The critical exponent $\delta$ is then the slope of
the best data fit to a straight line. Our estimate for the critical point
is $\omega=0.407$, where fitting the data into the straightest line gives a
slope $0.161$. Taking the interval $\omega \in \left [ 0.4065,0.4075
\right ]$, the data fits to slopes $0.1724$ and $0.1601$ respectively, which
gives an error estimate to $\delta=0.161(6)$. An interval for $\omega_c$
with range $\pm 10^{-3}$ overestimates the error on the exponent as $\omega
\in \left [ 0.406,0.4078  \right ]$ results in $\delta=0.161\pm0.01$. In all
the following estimates of the critical exponents we report the overestimated
error from data within range $\pm 10^{-3}$ of the driving parameter.

Next, we plot $n(t) \cdot t^\delta$ at different values of $\omega$
as a function of $t\Delta^{\nu_\parallel}$ and varying $\nu_\parallel$
to obtain the best collapse of the data into one universal scaling
function. $\nu_\parallel=1.7$ is the value that gives the best data
collapse plot shown in \cref{figclassical}c, whereas increments
of $0.05$ to $\nu_\parallel$ give clearly worse collapse plots
and thus the error on $\nu$ and $z$ is estimated as $0.03$. Using
$\nu_\parallel=1.70(3)$ and $\delta=0.161(6)$, we estimate the critical
exponent $\beta=\delta\cdot\nu_\parallel=0.27(2)$. Finally we plot $n(t)\cdot
t^\delta$ at different system sizes $L$ and $\omega=\omega_c$ as a function
of $t/L^z$  (here $d=1$ and $N=L$). The exponent $z$ is varied to optimize
the collapse. We find $z=1.6(3)$ and deduce the last exponent of the triplet
$\nu_\perp=\nu_\parallel/z=1.06(4)$. All the critical exponents agree with
the known DP value, see \cref{tab:exp}.

%///////////////////////////////////////////////////////////////////////////////
\begin{figure}[tb]
    \centering
    \includegraphics[width=\columnwidth]{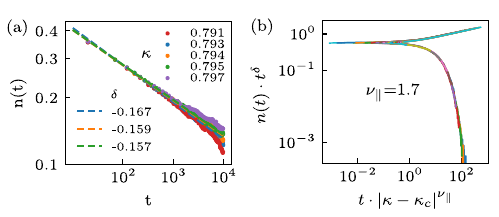}
    \caption{%
        \CaptionMark{The \CondSX model.} 
        We observe that the scaling behavior is consistent with the expected
        classical DP universality class. The data shown is for the system size
        $L=400$ and $\kappa_c=0.7940(5)$ is the estimated critical point. The
        critical exponents are then $\delta=0.159(3)$, $\nu_\parallel=1.70(3)$,
        and $\beta=0.27(1)$ (cf.\ \cref{tab:exp}).
        }
    \label{figclassical2}
\end{figure} 
%///////////////////////////////////////////////////////////////////////////////

\begin{figure}[tb]
    \centering
    \includegraphics[width=0.48\textwidth]{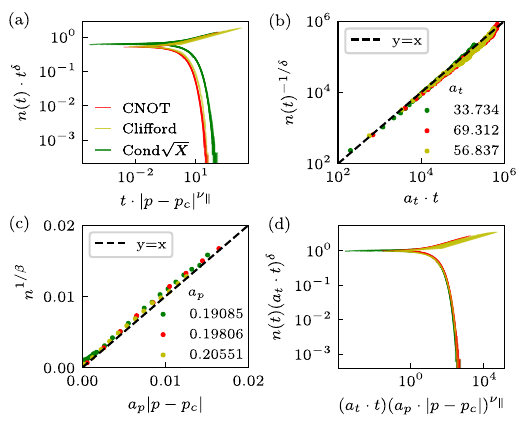}
    \caption{%
    \CaptionMark{Non-universal scaling factors.}
    \lbl{a}~Collapse plots of the considered models with different
	spreading gates, where $p \in \{\omega, \kappa\}$ is the driving
	parameter of the phase transition.
    \lbl{b,c}~Scaling of the simulation data from different models with
    appropriate factors $a_p$ and $a_t$ which encode non-universal microscopic
    differences specific for each model. 
    \lbl{d}~Scaled collapse plots which now all fall on top the universal
    scaling function of the DP class.
    }
    \label{non-universal}
\end{figure}

A similar analysis is shown in \cref{figclassical2} for the incoherent \CondSX
model, which also falls within the DP universality class, as well as for the
Clifford model discussed in the main text (\cref{Result2}). We conclude that
all these contact models share the same critical exponents, despite having
different activation spreading processes defined by the dynamical rules
illustrated in \cref{fig:model}. To prove that they indeed belong to the
same universality class, we should also show that the scaling functions $f$
coincide. Plotting $n(t)t^\delta$ as a function of $t\Delta^{\nu_\parallel}$
for the three models at the same decay rate $\gamma$ and system size $L$
does not show this result (\cref{non-universal}a). However, after scaling
the simulation data with appropriate non-universal factors that encode the
microscopic properties specific for each model, the scaling functions from
the different models indeed collapse to the same universal function, as shown
in \cref{non-universal}d. \cref{non-universal}b and \cref{non-universal}c
show the results obtained for the non-universal factors $a_p$ and $a_t$
respectively. We can estimate these factors by plotting the scaling relations
$n(t_s)=|a_p(p-p_c)|^\beta$ and $n(t)=a_t\, t^{-\delta}$ (which are
valid near the critical point) in the double logarithmic scales and extract
the different slopes corresponding to the data from different models.

%///////////////////////////////////////////////////////////////////////////////
%///////////////////////////////////////////////////////////////////////////////

%+++++++++++++++++++++++++++++++++++++++++++++++++++++++++++++++++++++++++++++++
\section{Longer interaction range}
\label{app:range}
%+++++++++++++++++++++++++++++++++++++++++++++++++++++++++++++++++++++++++++++++

\begin{figure}[tb]
    \centering
    \includegraphics[width=0.9\columnwidth]{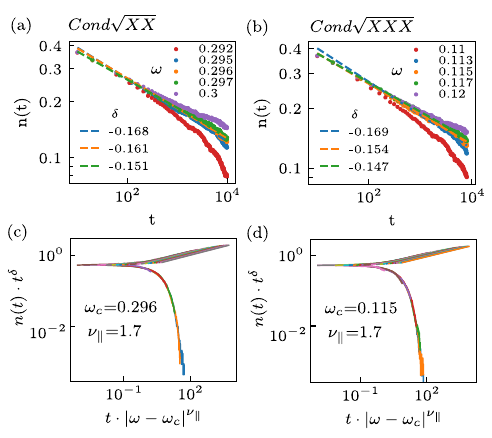}
    \caption{%
        \CaptionMark{Critical exponents for the \CondSXX and the \CondSXXX models.}
        \lbl{a}~We find the exponent $\delta=0.161(4)$ at the critical point
        $\omega_c=0.2960(5)$ for the \CondSXX model.
        \lbl{c}~From the data collapse, we estimate $\beta=0.27(1)$. 
        \lbl{b,d}~A similar analysis for the \CondSXXX model estimates
        the exponents $\delta=0.154(6)$ at $\omega_c=0.1150(5)$ and
        $\beta=0.26(2)$. In both models, the transition remains clearly
        within the DP universality class, only the transition point shift
        towards lower spreading rate, which is expected as further sites
        can be activated by these three and four-qubit operations.
    }
    \label{fignext}
\end{figure}

Here we discuss the effect of replacing the two-qubit \CondSX operation in the
Clifford model with a three-qubit \CondSXX operation or a four-qubit \CondSXXX
operation on the nature of the observed absorbing state phase transition. The
universality class can depend on the range of interactions. Models where
activation spreads over longer distances have been investigated, an example
is models with L\'{e}vy flight distributions~\cite{bouchaud1990anomalous}. The
universal properties of such processes are known to diverge from that of
DP continuously with the control parameter $\sigma$ which defines the shape
of the probability distribution $P(r) \sim 1/r^{d+\sigma}$ where $r$ is the
distance over which a random interaction can occur; for $\sigma=\infty$ one
gets back the DP case.  Contact processes with short-range interaction belong
to the DP universality class. In the Clifford models we define, the range of
interaction is still short relative to the system size, in particular, it is
bounded. So that the nature of the phase transition remains unchanged. This
is discussed in \cref{fignext} where the obtained critical exponents are
still sufficiently close to DP.

%///////////////////////////////////////////////////////////////////////////////
%///////////////////////////////////////////////////////////////////////////////

%+++++++++++++++++++++++++++++++++++++++++++++++++++++++++++++++++++++++++++++++
\section{Small system sizes}
\label{app:small}
%+++++++++++++++++++++++++++++++++++++++++++++++++++++++++++++++++++++++++++++++

\begin{figure*}[tb]
    \centering
    \includegraphics[width=0.90\textwidth]{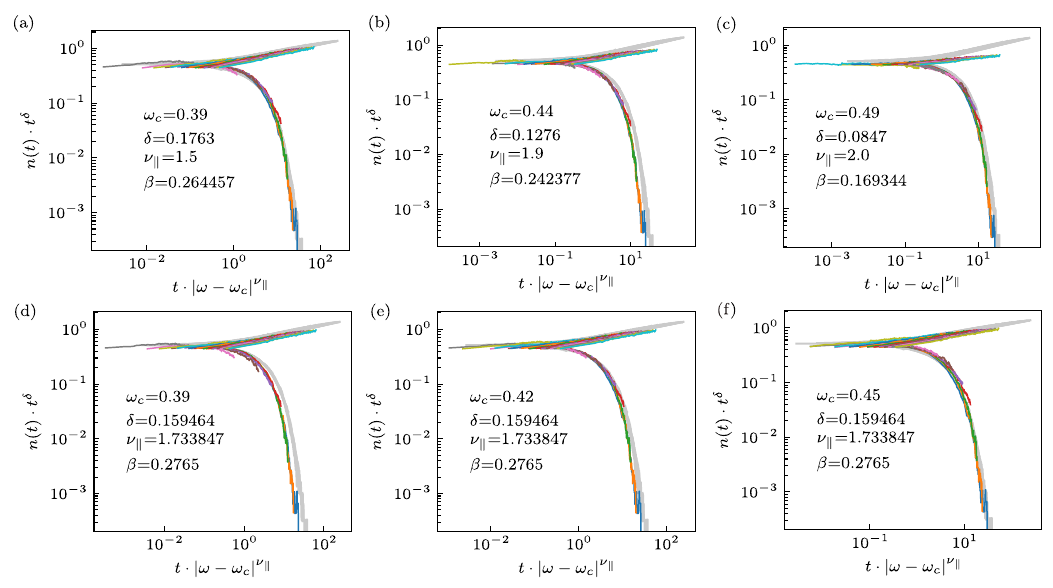}
    \caption{%
        \CaptionMark{Collapse plots for the \CNOT model at $L=20$.} 
        The large system data is plotted in gray as a background for reference.
        \lbl{a-c}~In the upper plots $\omega_c$ is varied in the range shown
        in \cref{Result7} and $\delta$ is the fitted exponent. We observe
        that the data from the L=20 simulation agrees with the thermodynamic
        results, but at a lower critical value $\omega_c \approx 0.39$
        which is less than $\omega_c =0.407$ obtained for $L=400$.
        \lbl{d-f}~For the lower collapse plots we fix the literature values
        for the DP exponents, $\delta=0.159$ and $\nu_{\parallel}=1.73$, and
        vary the critical point $\omega_c$. We observe good data collapses
        within a parameter range $0.39<\omega_c<0.45$. The best fit occurs
        around $\omega_c=0.42$ which slightly higher than the thermodynamic
        value $\omega_c =0.407$. 
    }
    \label{collCNOT}
\end{figure*}

In this section we investigate how susceptible the data collapse is to
variations of the critical exponents for small system sizes.  We start with
the results of the \CNOT model where the critical behavior is known from
larger system simulations. For small sizes, the transition point is hard to
identify (\cref{Result7}). Finite size effects dominate the long time regime,
we end up with a wide range of spreading rates and corresponding slopes
in the logarithmic scale where the decay appears linear. In the $L=20$
\CNOT model, as shown in \cref{Result7}, the range appears to be roughly
$0.4<\omega_c<0.5$ and $0.08<\delta<0.17$. \cref{collCNOT}a shows the data
collapsed with parameters within this range. The error is of the order
of the exponent which is expected for such small sizes where extracting
exponents from the data is impossible. However, we notice that the $L=20$
data collapses nicely on top of the $L=300$ data (plotted in gray) when we
the exponents are close to the expected DP values. This makes us conclude
that it is possible, in the case of the \CNOT model, to verify the nature
of the transition by scaling the simulation data using the known literature
values of the DP exponents. By varying the spreading coupling $\omega$,
we find an $\omega_c=0.42$ where the data shows a good collapse, as seen in
\cref{collCNOT}b. We remark that a range of values $\omega \in [0.39,0.45]$
give relatively good collapse plots, the expected $\omega_c=0.407$ is within
this interval.

\begin{figure*}[tb]
    \centering
    \includegraphics[width=0.90\textwidth]{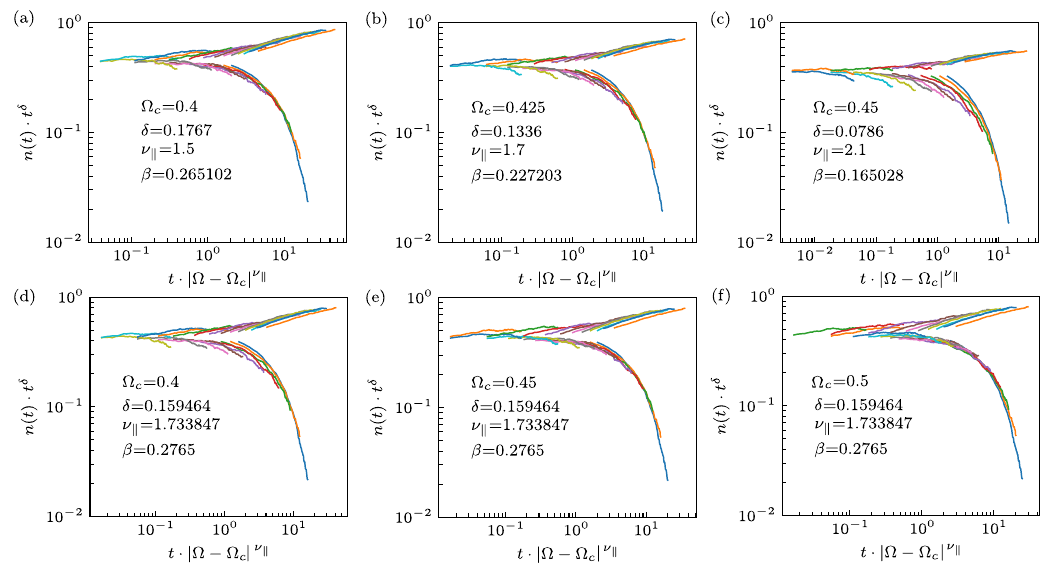}
    \caption{%
        \CaptionMark{Collapse plots for \CSX model at L=20.} 
        We use the data between $10<t<100$ time steps. Plots in \lbl{a-c}
        use fitted $\delta$ exponents while collapse plots in \lbl{d-f}
        use the DP exponents. Based on this data we cannot exclude a contact
        process transition within the DP universality class in the \CSX model
        where the critical point occurs for coupling $0.4<\Omega_c<0.45$.
        %\todo{Labels correct?}
    }
    \label{collCSX}
\end{figure*}

A similar analysis is done on the \CSX model with $L=20$. For this model,
the critical behavior is unknown. The collapse plots are shown in \cref{collCSX}, 
they appear significantly worse than the ones obtained for the \CNOT model.
However, we cannot exclude a DP transition in the range $0.4<\Omega_c<0.45$ 
as the upper left most plot and the lower middle plot have a relatively decent 
collapse quality considering the size limitation.

\clearpage

%///////////////////////////////////////////////////////////////////////////////
%///////////////////////////////////////////////////////////////////////////////

%///////////////////////////////////////////////////////////////////////////////
%///////////////////////////////////////////////////////////////////////////////

\end{appendix}

\end{document}